\documentclass[a4paper,fleqn]{cas-dc}
\usepackage{graphicx, float}
\usepackage{amsmath}
\usepackage{amsthm}  
\usepackage{mathrsfs}
\usepackage[utf8]{inputenc}
\usepackage{rotating}
\usepackage{multirow}
\usepackage{booktabs}
\usepackage{siunitx}
\usepackage{multirow}
\usepackage{tikz}

\newlength{\spanunit}
\newlength{\spanhalf}

\usepackage{caption}
\usepackage[position=top]{subcaption}
\usepackage{comment}
\usepackage{bbding}
\usepackage[export]{adjustbox}
\usepackage{standalone}
\usepackage{rotating}
\usepackage[T1]{fontenc} 
\usepackage{utfsym}

\usepackage[per-mode=symbol]{siunitx}
\usepackage[normalem]{ulem}
\usepackage{cancel}
\usepackage{pifont}
\usepackage[dvipsnames]{xcolor}  
\usepackage{hyperref}
\hypersetup{hidelinks}
\usepackage[nameinlink,noabbrev]{cleveref}
\crefname{equation}{equation}{equations}
\crefname{algorithm}{Algorithm}{Algorithms}
\Crefname{algorithm}{Algorithm}{Algorithms}
\usepackage{algorithm}      
\usepackage{algpseudocode} 
\usepackage{bm}
\usepackage{tikz}
\usetikzlibrary{decorations.pathreplacing}
\usepackage{breakcites}
\usetikzlibrary{shapes}
\usepackage{mathtools}
\definecolor{red2}{RGB}{192, 16, 16}
\definecolor{green2}{RGB}{48, 128, 48}
\definecolor{green3}{RGB}{0, 200, 0}
\definecolor{blue2}{RGB}{16, 80, 160}
\definecolor{blue3}{RGB}{0, 0, 255}
\definecolor{orange2}{RGB}{192, 128, 64}
\definecolor{yellow2}{RGB}{224, 192, 32}
\definecolor{purple2}{RGB}{96, 64, 160}
\definecolor{teal2}{RGB}{32, 112, 144}

\definecolor{drkg}{RGB}{105,105,105}
\definecolor{midg}{RGB}{169,169,169}
\definecolor{lgtg}{RGB}{220,220,220}
\definecolor{drkb}{RGB}{0,25,51}
\definecolor{lgtb}{RGB}{0,102,204}
\definecolor{lightg}{RGB}{160,160,160}
\definecolor{lightgr}{RGB}{30,120,30}

\usepackage[numbers, sort&compress]{natbib}
\def\tsc#1{\csdef{#1}{\textsc{\lowercase{#1}}\xspace}}
\tsc{WGM}
\tsc{QE}
\tsc{EP}
\tsc{PMS}
\tsc{BEC}
\tsc{DE}

\usepackage{tikz}

\newlength{\topspanheight}

\begin{document}

\let\WriteBookmarks\relax
\def\floatpagepagefraction{1}
\def\textpagefraction{.001}

\shorttitle{Soft material mechanical property degradation at high strain rates}

\shortauthors{V. Sanchez et~al.}

\title[mode = title]{Soft material mechanical property degradation at high strain rates} 

\author[1]{Victor Sanchez}[orcid=0009-0002-9609-6520]
\cormark[1]

\credit{Conceptualization of this study, Methodology, Software, Data curation, Writing - Original draft preparation} 

\author[2]{ Joseph Beckett}

\credit{Data curation, Writing - Original draft preparation} 

\author[2]{ Aryan Shah}

\credit{Data curation, Writing - Original draft preparation} 

\author[3]{ Sicong Wang}

\credit{Data curation, Writing - Original draft preparation} 

\author[3,4]{ Jin Yang}[orcid=0000-0002-5967-980X]

\credit{Writing - Original draft preparation, Writing - review \& editing, Funding Acquisition} 

\author[2]{ Jonathan B. Estrada}[orcid=0000-0003-1083-4597]

\credit{Writing - Original draft preparation, Writing - review \& editing, Funding Acquisition} 

\author[1]{ Mauro Rodriguez Jr.}[orcid=0000-0003-0545-0265]
\cormark[2]
\credit{Conceptualization of this study, Writing - original draft, Writing - review \& editing, Funding Acquisition, Project Administration, Supervision}

\affiliation[1]{organization={School of Engineering, Brown University}, 
                city={Providence},
                postcode={02912}, 
                state={RI},
                country={USA}}

\affiliation[2]{organization={Department of Mechanical Engineering, University of Michigan}, 
                city={Ann Arbor},
                postcode={48109}, 
                state={MI},
                country={USA}}
                
\affiliation[3]{organization={Department of Aerospace Engineering and Engineering Mechanics, The University of Texas at Austin}, 
                city={Austin},
                postcode={78712}, 
                state={TX},
                country={USA}}

\affiliation[4]{organization={Texas Materials Institute, The University of Texas at Austin}, 
                city={Austin},
                postcode={78712}, 
                state={TX},
                country={USA}}


\cortext[cor1]{Corresponding author}
\cortext[cor2]{Principal corresponding author}

\begin{abstract}
The rheometry of soft, tissue-like materials at ultra-high strain rates is critical to biomechanics, materials science, and medical engineering.
One approach for characterizing high-rate material response is Inertial Microcavitation Rheometry (IMR), which infers viscoelastic constitutive behavior from the measured radius--time history of a laser-induced cavitation bubble.
At these rates, materials may exhibit history-dependent behavior, including damage, fracture, and plasticity.
However, conventional IMR analyses typically assume constant material properties throughout the cavitation event, limiting their ability to capture such changes.
We present an IMR framework for identifying effective time-varying material properties from measured cavitation radius histories.
The framework allows the shear modulus, viscosity, or both to evolve during the event using normalized sigmoid histories.
Synthetic tests recover prescribed constant and time-varying material histories.
Application to polyacrylamide and gelatin datasets yields effective material-property histories across formulations and experimental conditions.
Viscosity decreased across the datasets, while the extent of shear modulus reduction varied with material and formulation. 
When both properties decreased, their reductions could occur at different stages of cavitation.
This distinction in mechanical property degradation depends on both the property and its evolution during high-strain events.
\end{abstract}

\begin{highlights}
    \item Time-varying rheometry to detect degradation-like material behavior from inertial microcavitation events.
    \item Distinguish between constant and softening material property behavior using measured radius histories.
    \item Mapped effective degradation-like material property evolution across polyacrylamide and gelatin experiments.
\end{highlights}


\maketitle

\section{Introduction}
Soft materials can experience extreme stress-loading environments in biomechanics, medicine, and engineering applications.
Examples include blast- and impact-induced traumatic brain injury~\cite{Eyal2016, Finan2024, Lang2021, Estrada2021}, therapeutic ultrasound~\cite{Maxwell2013, Haskell2023,Vlaisavljevich2014,Vlaisavljevich2015}, and high-rate biomedical device interactions~\cite{Roberts2005, Dougan2022}.
These materials may also show nonlinear elasticity, rate-dependent dissipation, and cavitation-induced damage~\cite{Cohen2015,Cohen2010,Sengul2020,Mullins1969,Diani2009,GentLindley1959,Ball1982,Fond2001,Barney2020,kolluri2025determination,Wang2026}, which complicate constitutive model identification.
Accurately characterizing the viscoelastic response at strain rates spanning $\SIrange{E3}{E8}{1\per\second}$ remains challenging, as traditional rheometry techniques are typically limited to quasi-static or moderate-rate loading~\cite{Chen2011, Chen2016}.

In addition to their rate dependence, the responses of soft materials may depend on the loading history.
In rubber-like materials, for example, prior deformation can produce stress softening associated with the Mullins effect~\cite{Mullins1969,Diani2009}.
In continuum damage mechanics, damage can be represented using internal state variables within an irreversible-thermodynamic framework, with its mechanical effects expressed through effective stress or strain measures~\cite{SimoJu1987}.
These descriptions distinguish observable softening of the mechanical response from damage represented as an irreversible internal process.

Cavitation rheometry provides a means of probing these responses under high-strain-rate and finite-strain loading~\cite{Zimberlin2007, Barney2020, Dougan2024}.
The Inertial Microcavitation Rheometry (IMR) technique uses the collapse of laser-induced cavitation (LIC) bubbles to generate ultra-high-strain-rate deformations in a surrounding medium \cite{Estrada2018, Yang2019, Yang2020, Yang2021b, Buyukozturk2022}.
The resulting bubble radius--time history depends on the surrounding material's viscoelastic behavior~\cite{Gaudron2015, Warnez2015, Barajas2017, Abu2022, Tzoumaka2023}, enabling constitutive parameters to be inferred by fitting simulated histories to experimental observations.
The IMR technique has been validated for agarose~\cite{Yang2021b}, fibrin, gelatin~\cite{mcghee2023high,bremer2024ballistic}, polyacrylamide~\cite{Estrada2018,Yang2020}, and brain tissue~\cite{Dougan2022, Dougan2024}, as well as in direct comparisons of acoustically and laser-induced bubbles~\cite{Wilson2019, Abeid2024}.

Recent studies have improved the statistical rigor, efficiency, and interpretability of IMR results.
For example, \citet{Spratt2021} used ensemble-based data assimilation to infer viscoelastic properties under uncertainty, revealing apparent viscosity degradation near collapse that suggested limitations of the assumed constitutive model.
\citet{Zhu2025} and \citet{yang2024estimating} used a collapse-time energy budget framework to efficiently analyze multiple experimental trials.
A previous hierarchical Bayesian model-selection study identified the neo-Hookean Kelvin--Voigt model as one of the most consistently supported descriptions across comparable cavitation datasets while screening out non-credible best-fit models~\cite{Sanchez2026}.
\citet{Chu2026} used moving-window data assimilation to recover time-resolved apparent shear modulus and viscosity.
However, their window locations were prescribed uniformly and not informed by features in the radius history itself.
Thus, it is still unclear which time-varying behaviors are necessary to explain the dynamics.

A current challenge is determining how time-varying mechanical material properties should be interpreted and whether such changes reflect irreversible mechanisms.
Standard IMR analyses return an effective parameter set over a prescribed time window.
However, soft materials may undergo cavitation-induced damage or fracture during this window under sufficiently strong loading conditions~\cite{Hasan2021,RaayaiArdakani2019,Dougan2022}.
Damage-aware cavitation studies have used material failure criteria, tissue susceptibility, and local mechanical fields to find conditions under which bubble collapse produces damage in soft materials~\cite{Movahed2016,Mancia2017,Mancia2019,Mancia2021}.
\citet{Wang2026} extended this technique to anisotropic PVA hydrogels and chicken breast tissue using a fiber-stretch-based damage model to estimate effective directional moduli and critical stretches for damage initiation.
Collectively, these approaches prescribe a damage model, apply a failure criterion, and/or compute local fields to predict the onset or extent of cavitation-induced damage.
It therefore remains unclear whether the measured radius dynamics are adequately described by constant or time-varying material properties within an assumed constitutive model and/or are influenced by other physics.

We present an event-based IMR technique for identifying effective time-varying material properties from measured cavitation radius histories.
The technique allows the shear modulus, viscosity, or both to evolve during the bubble-dynamics event.
The neo-Hookean Kelvin--Voigt constitutive model is adopted as the baseline~\cite{Sanchez2026}.
The technique identifies effective temporal reductions in material resistance that are consistent with mechanical property degradation.
The paper is organized as follows.
In~\cref{sec:problemformulation}, the equations of motion, event windows, and material-history parameterization are introduced.
The optimization procedure and material-history analysis are detailed in~\cref{sec:optimization}.
In~\cref{sec:results}, the method is validated using synthetic cases and applied to LIC experimental trials, with the resulting trends discussed in context.
The main contributions are summarized in~\cref{sec:conclusions}.

\section{Problem formulation\label{sec:problemformulation}}
\subsection{Equations of motion\label{sec:equation_of_motion}}
Spherical bubble dynamics in an incompressible viscoelastic medium with weakly compressible far-field acoustics can be described by the Keller--Miksis equation~\cite{Keller1980,Prosperetti1986a}:
\begin{equation}
    \begin{aligned}
        \left(1-\frac{\dot{R}}{c}\right) R \ddot{R} &+ \frac{3}{2}\left(1-\frac{\dot{R}}{3 c}\right) \dot{R}^{2} = \\ &\left(1+\frac{\dot{R}}{c} + \frac{R}{c}\frac{d}{dt}\right)\frac{\Delta p}{\rho},
    \end{aligned}
    \label{eq:KM_R}
\end{equation}
where the dot operators denote time derivatives, $R$ the bubble radius, $c$ the speed of sound in the material surroundings, $\rho$ the surrounding material density, and $\Delta p$ the pressure difference between the inside and outside of the bubble.
The equation is nondimensionalized using the maximum bubble radius $R_{\max}$, surrounding material density $\rho$, ambient pressure $p_\infty$, time scale $t_c = R_{\max}/\sqrt{p_\infty/\rho}$ and collapse speed $v_c = R_{\max}/t_c$.
For example, $R^*=R/R_{\max}$ is the nondimensional bubble radius, $R_{\mathrm{eq}}^*=R_{\mathrm{eq}}/R_{\max}$ the nondimensional equilibrium radius, and $t^*=t/t_{\mathrm{c}}$ the nondimensional time.
Hereafter, superscript asterisks are dropped.
The pressure difference is
\begin{equation}
    \Delta p = p_{\text{b}}-1-\frac{1}{\text{We} R}+S,
    \label{eq:KM_p}
\end{equation}  
where $p_{\mathrm{b}}$ is the total bubble pressure accounting for heat transfer within the bubble and surrounding medium and interfacial mass transfer~\cite{Estrada2018}, $\text{We}= \rho v_c^2 R_{\max}/(2\gamma)$ the Weber number with surface tension $\gamma$, and $S$ the stress integral.
For a neo-Hookean Kelvin--Voigt constitutive model, the stress integral is
\begin{equation}
    S=\frac{1}{2\text{Ca}}\left[4\lambda^{-1}+\lambda^{-4}-5\right] -\frac{4}{\text{Re}}\frac{\dot{R}}{R},
    \label{eq:stress_integral}
\end{equation}
where $\text{Ca}=p_\infty/G$ is the Cauchy number with shear modulus $G$, $\lambda=R/R_{\text{eq}}$ the stretch ratio associated with the spherical bubble deformation, $R_{\text{eq}}$ the equilibrium bubble radius, and $\text{Re}=\rho v_c R_{\max}/\mu$ the Reynolds number with viscosity $\mu$.
The maximum stretch ratio is $\Lambda=R_{\max}/R_{\text{eq}}$.
The shear modulus and viscosity, denoted collectively by $\theta(t)\in\{G(t),\mu(t)\}$, are allowed to vary in time to assess whether effective mechanical property degradation improves agreement with the measured radius history.
Such degradations do not identify material damage.
The radial Hencky strain at the bubble wall is $\varepsilon=-2\ln\lambda$, with corresponding dimensionless strain rate $\dot{\varepsilon}=-2\dot{R}/R$.

\subsection{Experimental methods and data}
\label{sec:experiments}

\begin{table*}[!t]
\centering
\caption{Materials, concentrations, sample sizes, and bubble-geometry statistics for experimental laser-induced cavitation datasets.
PA concentrations are listed as AAM/BIS wt\%.
Bubble-geometry values are mean $\pm$ standard deviation across experimental trials.}
\label{tab:materials}

\small
\setlength{\tabcolsep}{7pt}
\renewcommand{\arraystretch}{1.1}

\begin{tabular}{@{}llcccc@{}}
\toprule
Dataset & Material & Concentration [wt\%]& $N$
& $R_{\max}$ [\si{\micro\meter}] & $\Lambda$ \\
\midrule
PA-3-015  & PA & 3.0/0.150 & 37
& $359.9 \pm 69.8$ & $9.06 \pm 0.33$ \\
PA-3-0225 & PA & 3.0/0.225 & 32
& $407.7 \pm 59.7$ & $8.94 \pm 0.28$ \\
PA-3-030  & PA & 3.0/0.300 & 32
& $439.8 \pm 46.9$ & $8.74 \pm 0.31$ \\
PA-5-030  & PA & 5.0/0.300 & 31
& $325.1 \pm 63.4$ & $7.90 \pm 0.32$ \\
KG-5      & Knox gelatin & 5.0 & 46
& $419.0 \pm 9.0$ & $7.78 \pm 0.23$ \\
KG-10     & Knox gelatin & 10.0 & 76
& $349.0 \pm 41.2$ & $5.20 \pm 0.18$ \\
PG-5-UM   & Porcine gelatin & 5.0 & 87
& $314.6 \pm 46.2$ & $7.43 \pm 0.15$ \\
PG-10-UM  & Porcine gelatin & 10.0 & 86
& $291.7 \pm 52.7$ & $5.07 \pm 0.14$ \\
PG-5-UT   & Porcine gelatin & 4.94 & 36
& $339.9 \pm 35.6$ & $7.85 \pm 0.39$ \\
PG-10-UT  & Porcine gelatin & 9.75 & 39
& $279.4 \pm 25.6$ & $5.40 \pm 0.16$ \\
\bottomrule
\end{tabular}
\end{table*}

\begin{figure}[!htb]
    \centering
    \includegraphics[width=\columnwidth]{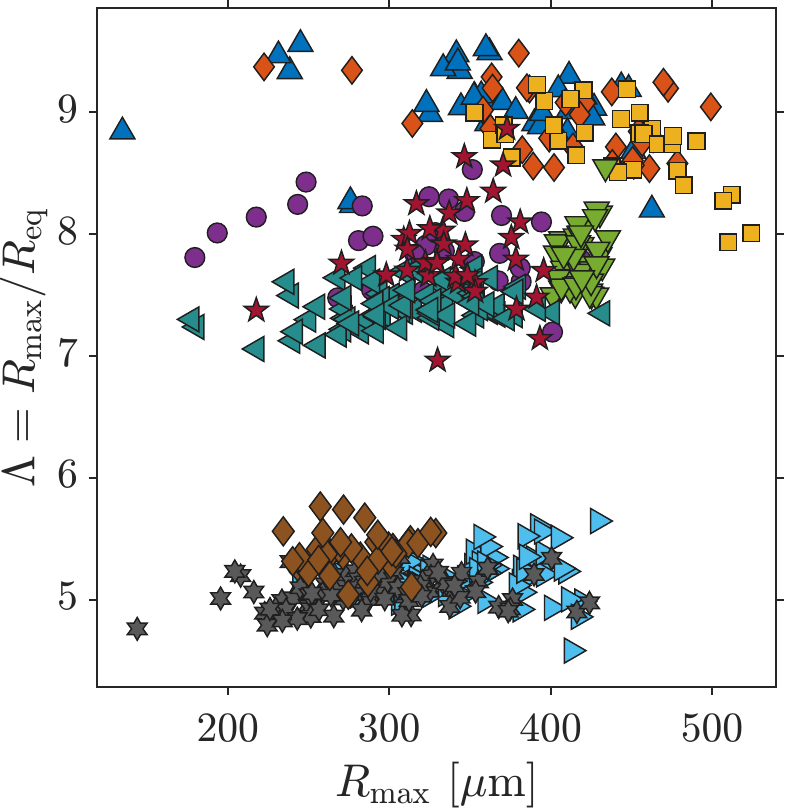}
    \caption{Maximum stretch ratio, $\Lambda=R_{\max}/R_{\mathrm{eq}}$, versus maximum bubble radius, $R_{\max}$, for 502 LIC trials across ten datasets.
    Each marker is one trial.
    Marker colors and shapes identify the datasets as follows.
    The PA datasets are PA-3-015 (blue upward triangle), PA-3-0225 (orange diamond), PA-3-030 (yellow square), and PA-5-030 (purple circle).
    The Knox gelatin datasets are KG-5 (green downward triangle) and KG-10 (cyan right-facing triangle).
    The porcine gelatin datasets are PG-5-UM (teal left-facing triangle), PG-10-UM (gray six-pointed star), PG-5-UT (red five-pointed star), and PG-10-UT (brown diamond).}
    \label{fig:lambda_vs_Rmax}
\end{figure}

The experimental database consists of 502 LIC trials across ten datasets collected at the University of Michigan (UM) and The University of Texas at Austin (UT), including four chemically cross-linked polyacrylamide (PA) and six gelatin datasets.
\Cref{tab:materials} summarizes the material compositions, trial counts, and bubble geometry by dataset, and \cref{fig:lambda_vs_Rmax} shows the corresponding trial-level distributions of $R_{\max}$ and $\Lambda$.

The UM experimental trials include polyacrylamide formulations with acrylamide/bisacrylamide concentrations of 3.0/0.150, 3.0/0.225, 3.0/0.300, and 5.0/0.300~wt\%, labeled PA-3-015, PA-3-0225, PA-3-030, and PA-5-030, respectively.
The UM gelatin datasets consist of Knox gelatin at 5 and 10~wt\% (KG-5 and KG-10) and porcine gelatin at 5 and 10~wt\% (PG-5-UM and PG-10-UM).
The UT datasets consist of porcine gelatin at 5 and 10~\% w/v (PG-5-UT and PG-10-UT). 
Across all datasets, the mean maximum bubble radii range from approximately $\SIrange{279}{440}{\micro\meter}$, while the mean maximum stretch ratios range from $5.07$ to $9.06$.

Microcavitation bubbles in both experimental sets were generated using a frequency-doubled, Q-switched Nd:YAG laser running at $\SI{532}{\nano\meter}$ with pulse durations of 3--5~ns. 
Bubble dynamics were recorded at $10^6$ frames per second using a Shimadzu HPV-X2 high-speed camera with synchronized collimated backlighting. 
Bubble radii were found through intensity-based thresholding and subsequent circle fitting, following the procedures described by \citet{Estrada2018} and \citet{wang2026imrfit}.

Minor geometry differences distinguished the two institutional experimental setups. 
UM bubble nucleation sites were spaced away from walls of vertical 5~mL cuvettes and previous cavitation sites to minimize boundary effects and interactions with previously cavitated regions. 
A diffraction-limited $10\times$ spot focusing objective with numerical aperture (NA) of 0.25 collected the beam for inducing cavitation. 
At UT, a $10\times$/0.3 NA objective focused the laser approximately $\SI{742}{\micro\meter}$ above the bottom of a petri dish situated on a microscope stage.
Stand-off distances were calculated by dividing the distance from the bubble center to each boundary by the maximum bubble radius. 
For PG-5-UT and PG-10-UT, respectively, the stand-off distances to the dish bottom were approximately 2.2 and 2.7, and those to the upper free surface exceeded 6.6 and 8.1.

For the UM PA datasets, 40\% w/v acrylamide, 2\% w/v $N,N'$-methylenebisacrylamide, and $N$, $N$, $N’$, $N’$-tetramethylethylenediamine (TEMED) were obtained from Bio-Rad Laboratories, and ammonium persulfate (APS) from MilliporeSigma.
Following \citet{Tse2010}, the stock solutions were diluted with deionized water to obtain the desired compositions. 
Each precursor was vortexed for $\SI{15}{\second}$ and vacuum-degassed for $\SI{15}{\minute}$. 
TEMED and 10\% w/v aqueous APS solution were then added at 0.1\% and 1\% of the precursor volume, respectively. 
After another $\SI{15}{\second}$ of mixing, $\SI{3.5}{\milli\liter}$ aliquots were transferred to $\SI{4.5}{\milli\liter}$ polystyrene cuvettes, sealed with Parafilm, and allowed to polymerize for at least $\SI{24}{\hour}$. 
Specimens were tested within one week of casting.

The UM Knox gelatin was dissolved in deionized water at $\SI{60}{\degreeCelsius}$ for $\SI{2}{\hour}$ to prepare 5\% and 10\% w/w mixtures. 
During heating, the mixtures were covered to limit evaporation and stirred slowly with a magnetic stir bar to keep a uniform temperature. 
The mixtures were transferred to cuvettes, covered with Parafilm, and allowed to gel for $\SI{24}{\hour}$. 
Specimens were tested within $\SI{48}{\hour}$ of preparation.

For the porcine gelatin experimental trials, UM and UT dissolved the same porcine gelatin powder (Type A, gel strength $\sim$300, MilliporeSigma) in deionized water at $\SI{80}{\degreeCelsius}$.
At UM, the 5\% and 10\% w/w mixtures were heated for $\SI{2}{\hour}$, using the same covering, stirring, casting, gelation, and testing procedures as for Knox gelatin.
At UT, 2.25 and 4.5~g of gelatin powder were mixed with deionized water in 50~mL centrifuge tubes to a final volume of 45~mL, yielding 5\% and 10\% w/v solutions.
The tubes were sealed to limit evaporation and heated for $\SI{60}{\minute}$.
The solutions were poured into glass-bottom Petri dishes to form cylindrical specimens approximately $\SI{10}{\milli\meter}$ in radius and more than $\SI{3}{\milli\meter}$ thick.
The dishes were covered with plastic lids, sealed with Parafilm, and refrigerated at $\SI{4}{\degreeCelsius}$ for at least $\SI{8}{\hour}$ to allow gelation.
Before testing, the sealed dishes were held at room temperature for at least $\SI{60}{\minute}$.
The UT specimens were tested within $\SI{24}{\hour}$ of preparation.
The UT hydrogel densities were estimated using an ideal mixing model with additive volumes~\cite{bailey2020viscoelastic}, with water and dry gelatin densities of $\SI{1.00}{\gram\per\centi\meter\cubed}$ and $\SI{1.35}{\gram\per\centi\meter\cubed}$, respectively.
The estimated hydrogel densities were $\SI{1.013}{\gram\per\centi\meter\cubed}$ for PG-5-UT and $\SI{1.026}{\gram\per\centi\meter\cubed}$ for PG-10-UT, corresponding to gelatin mass fractions of 4.94\% and 9.75\%, respectively.

\subsection{Event and history parameterization}

\begin{figure}[!htb]
    \centering
    \includegraphics[width=1.0\linewidth]{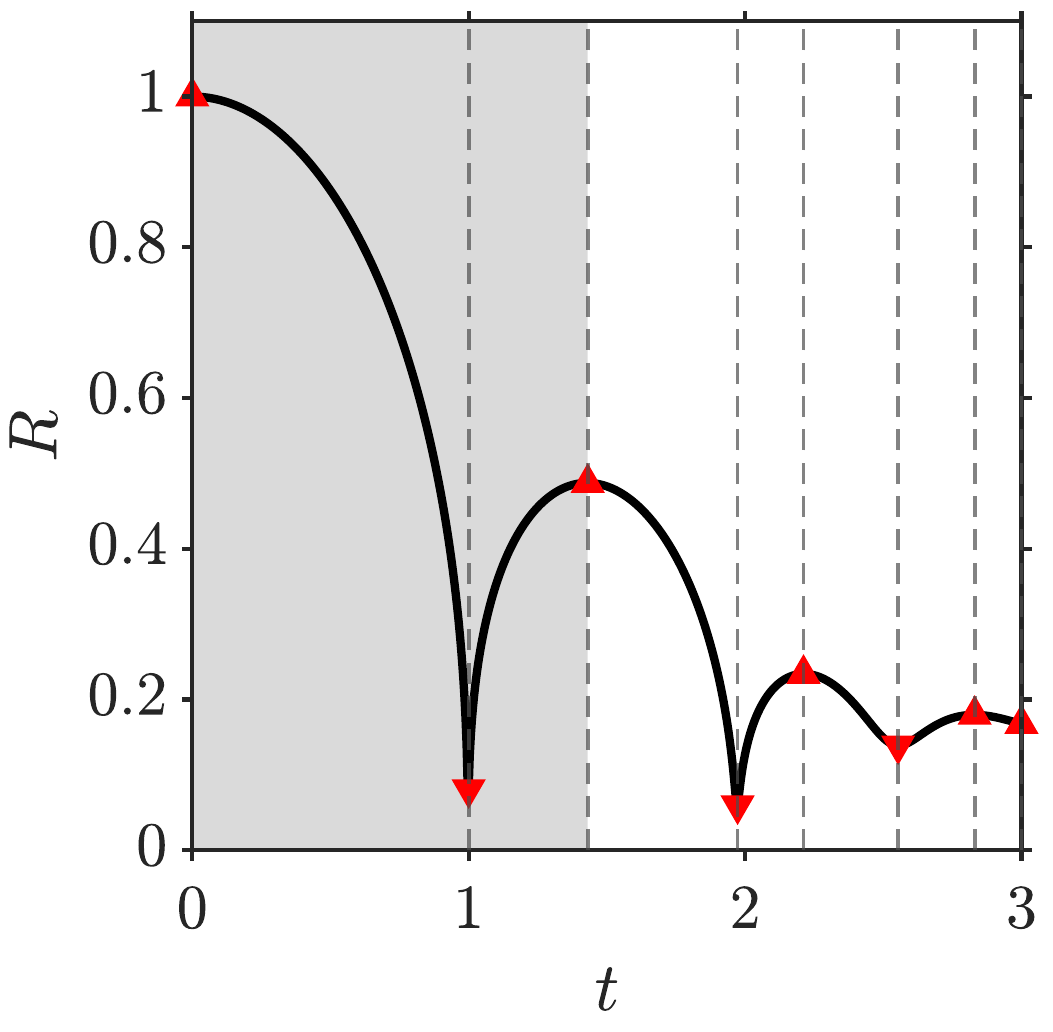}
    \caption{Representative radius history showing detected events $e_j$ (red triangles) and event-to-event windows $\mathcal{W}_m$ (consecutive vertical dashed lines). 
    The shaded region is the early-time interval $\mathcal{I}^0=\mathcal{W}_1\cup\mathcal{W}_2$ used to initialize the constant-property baseline.}
    \label{fig:Rt_ekw_method}
\end{figure}

\Cref{fig:Rt_ekw_method} shows a representative bubble radius history starting from the maximum bubble expansion radius. 
The model is initialized at the maximum bubble radius, which provides a clearly identifiable initial condition from the measured radius history~\cite{Estrada2018}.
The measured radius history is denoted by
\begin{equation*}
    \mathcal{D}=\left\{(t_n,R_n^{\text{obs}})\right\}_{n=1}^{N},
\end{equation*}
where $t_n$ is the $n$-th observation time, $R_n^{\mathrm{obs}}=R^{\mathrm{obs}}(t_n)$ the corresponding observed bubble radius, and $N$ the total number of observations.
The event set $\mathcal{E}$ contains the boundary times and detected local extrema of the observed radius history:
\begin{equation}
    \mathcal{E}=\{e_1=t_1,e_J=t_N\}\cup \left\{e_j:\frac{d R^{\text{obs}}}{dt}\bigg|_{t=e_j}=0\right\}_{j=2}^{J-1},
    \label{eq:event_set}
\end{equation}
where individual local events are $e\in\mathcal{E}$, $j$ indexes the events, and $J<N$ is the total number of detected events.
Consecutive events partition $\mathcal{D}$ into the set of $M=J-1$ event-to-event windows such that
\begin{equation}
    \mathcal{W}_m=\left\{n\in\{1,\ldots,N\}:e_m\leq t_n\leq e_{m+1}\right\},
    \label{eq:event_windows}
\end{equation}
where $m=1,\ldots,M$ (see dashed vertical lines).
The experimental analysis targets the interval through the third collapse to focus on the early high-amplitude cycles, where dominant property changes can be found before the bubble approaches equilibrium.

Over the event interval $[e_1,e_J]$, the evolution of each material property, $\theta\in\{G,\mu\}$, is represented by a normalized sigmoid history,
\begin{equation}
    \theta(t;a_\theta,r_\theta,t_{\mathrm{tr},\theta},w_\theta)=a_\theta\left[1+(r_\theta-1)\phi_\theta(t)\right],
    \label{eq:fitted_shape_history}
\end{equation}
where $a_\theta=\theta(e_1)>0$ is the initial value, $r_\theta=\theta(e_J)/\theta(e_1)$ the final-to-initial value ratio, and
\begin{equation}
    \phi_\theta(t)=\frac{\sigma_\theta(t)-\sigma_\theta(e_1)}{\sigma_\theta(e_J)-\sigma_\theta(e_1)},
    \label{eq:shape_basis}
\end{equation}
with
\begin{equation}
    \sigma_\theta(t)=\frac{1}{1+\exp\left[-(t-t_{\mathrm{tr},\theta})/w_\theta\right]}.
    \label{eq:logistic}
\end{equation}
$t_{\mathrm{tr},\theta}$ is the transition location and $w_\theta>0$ the width.
The normalization in \cref{eq:shape_basis} ensures that $\phi_\theta(e_1)=0$ and $\phi_\theta(e_J)=1$.
The degradation compared to the initial property value is $1-r_\theta$.

The history parameters control the magnitude and timing of property changes.
The constraint $0<r_\theta\leq1$ permits constant or decreasing histories, with $r_\theta=1$ giving constant behavior.
This focuses the analysis on whether an effective loss of material resistance can account for discrepancies in the constant-property description.
For a transition located within the analyzed interval, the normalized sigmoid approaches a step as $w_\theta\to0$ and a linear history as $w_\theta\to\infty$ with fixed transition location.
The shear modulus and viscosity are constrained within broad ranges encompassing those used in~\cite{Sanchez2026}:
\begin{subequations}
    \label{eq:parameter_bounds}
    \begin{align}
        \SI{E2}{\pascal} &\leq G(t) \leq \SI{E6}{\pascal},\\
        \SI{E-4}{\pascal\cdot\second} &\leq \mu(t) \leq \SI{10}{\pascal\cdot\second}.
    \end{align}
\end{subequations}

\section{Optimization problem setup\label{sec:optimization}}

\subsection{Forward simulations\label{sec:forwardsims}}
The forward simulations solve~\cref{eq:KM_R} using the IMR solver~\cite{Estrada2018,Sanchez2026,Zhu2025,Remillard2026Instability} with the material histories defined in~\cref{eq:fitted_shape_history}.
The material histories are parameterized by
$\mathbf{z}=[\mathbf{z}_G,\mathbf{z}_\mu]$, where
\begin{equation}
    \mathbf{z}_\theta=\left[\log a_\theta,r_\theta,t_{\mathrm{tr},\theta},w_\theta\right].
    \label{eq:optimization_variables}
\end{equation}
For the paired histories $\boldsymbol{\theta}(t;\mathbf{z})=[G(t;\mathbf{z}_G),\mu(t;\mathbf{z}_\mu)]$, the simulated radius is
\begin{equation}
    R^{\mathrm{sim}}(t;\boldsymbol{\theta})
    =\mathcal{F}\left(t;\boldsymbol{\theta}(t;\mathbf{z})\right),
    \label{eq:forward_operator}
\end{equation}
where $\mathcal{F}$ denotes the IMR forward solver (\Cref{eq:KM_R}).

\subsection{Rheometry\label{sec:objective}}
\Cref{alg:candidate_selection} summarizes the fitting procedure.
Initial material-property estimates are obtained from a successively refined constant-property (i.e., $r_G=r_\mu=1$) grid search over the early-time interval:
\begin{equation}
\label{eq:baseline_fit}
    \boldsymbol{\theta}^{0}=\arg\min_{\boldsymbol{\theta}}\sum_{n\in\mathcal{I}^{0}}\left[R^{\mathrm{sim}}(t_n;\boldsymbol{\theta})-R^{\mathrm{obs}}(t_n)\right]^2,
\end{equation}
where $\mathcal{I}^{0}=\mathcal{W}_1\cup\mathcal{W}_2=\{n:e_1\leq t_n\leq e_3\}$ is the shaded early-time interval spanning the first two event-to-event windows (see~\cref{fig:Rt_ekw_method}).

Starting from the initial estimates $\boldsymbol{\theta}^{0}$, the time-varying material histories are fitted by minimizing an objective function,
\begin{equation}
    \mathbf{z}^\star=\arg\min_{\mathbf{z}}\mathcal{J}\left(\boldsymbol{\theta}(t;\mathbf{z})\right),
    \label{eq:inverse_problem}
\end{equation}
where the star superscript denotes optimized quantities.
The objective function combines the global and window-balanced relative errors,
\begin{equation}
    \mathcal{J}(\boldsymbol{\theta})=\left(1-\omega\right)E^{\text{glob}}(\boldsymbol{\theta})+\omega E^{\text{win}}(\boldsymbol{\theta}),
    \label{eq:objective}
\end{equation}
where $\omega=0.5$ assigns equal weight to the global and window-balanced relative errors.

The global relative radius error is 
\begin{equation}
    E^{\text{glob}}(\boldsymbol{\theta})=\frac{\left\|\Delta\mathbf{R}(\boldsymbol{\theta})\right\|}{\left\|\mathbf{R}^{\text{obs}}\right\|},
    \label{eq:global_error}
\end{equation}
where 
\begin{equation}
    \Delta\mathbf{R}(\boldsymbol{\theta})= \left[ R^{\text{sim}}(t_n;\boldsymbol{\theta})-R^{\text{obs}}(t_n) \right]_{n=1}^{N},
\end{equation}
is the residual vector and $\mathbf{R}^{\text{obs}}= \left[ R^{\text{obs}}(t_n) \right]_{n=1}^{N}$ the observed radius vector.
The window-balanced error is the weighted mean relative error over event windows:
\begin{equation}
    E^{\text{win}}(\boldsymbol{\theta})=\frac{\sum_{m=1}^{J-1}N_m E_m(\boldsymbol{\theta})}{\sum_{m=1}^{J-1}N_m},
    \label{eq:window_balanced_error}
\end{equation}
where $N_m$ is the number of observations in window $\mathcal{W}_m$ and
\begin{equation}
    E_{m}(\boldsymbol{\theta})=\frac{\left\|\Delta \mathbf{R}_{m}(\boldsymbol{\theta})\right\|}{\left\|\mathbf{R}^{\text{obs}}_m\right\|},
    \label{eq:window_error}
\end{equation}
is the corresponding window relative error, with $\Delta \mathbf{R}_{m}$ and $\mathbf{R}^{\text{obs}}_m$ denoting the residual and observed-radius vectors in $\mathcal{W}_m$, respectively.

The optimization uses MATLAB's \texttt{surrogateopt}, followed by local refinement using \texttt{fmincon}, to minimize~\cref{eq:objective}.
The solution with the lowest objective value is retained as the best-fitting recovered history, giving
\begin{equation}
    \boldsymbol{\theta}^\star(t)=\boldsymbol{\theta}(t;\mathbf{z}^\star).
    \label{eq:selected_candidate}
\end{equation}

\begin{algorithm}[!t]
\caption{Material-history optimization}
\label{alg:candidate_selection}
\begin{algorithmic}[1]
\Require Measured radius data $\mathcal{D}$ and forward operator $\mathcal{F}$
\Ensure Recovered material history $\boldsymbol{\theta}^{\star}(t)$
\State Define $\mathcal{E}$ and $\{\mathcal{W}_m\}_{m=1}^{M}$ using~\cref{eq:event_set,eq:event_windows}
\State Obtain $\boldsymbol{\theta}^{0}$ using~\cref{eq:baseline_fit}
\State Construct $\boldsymbol{\theta}(t;\mathbf{z})$ using~\cref{eq:fitted_shape_history,eq:optimization_variables}
\State Use \texttt{surrogateopt} with $\boldsymbol{\theta}^{0}$ to identify $K$ starting points $\{\mathbf{z}^{(k)}\}_{k=1}^{K}$ for local refinement
\For{$k=1,\ldots,K$}
    \State Refine $\mathbf{z}^{(k)}$ using \texttt{fmincon} and~\cref{eq:objective}
\EndFor
\State For experimental data, also minimize~\cref{eq:objective} with $r_G=r_\mu=1$
\State Retain the lowest-objective solution $\mathbf{z}^{\star}$
\State Return $\boldsymbol{\theta}^{\star}(t)$ using~\cref{eq:selected_candidate}
\end{algorithmic}
\end{algorithm}

For synthetic validation, prescribed and recovered histories are compared using the relative recovery error
\begin{equation}
    E_{\theta}=\frac{\left\|\theta^{\star}-\theta^{\mathrm{true}}\right\|}{\left\|\theta^{\mathrm{true}}\right\|}.
    \label{eq:synthetic_recovery_error}
\end{equation}
For experimental comparisons, recovered histories are normalized by their fitted initial values, $\widehat{\theta}(t)=\theta^{\star}(t)/\theta^{\star}(e_1)$.
Dataset-level mean normalized histories, $\langle\widehat{\theta}\rangle$, and their pointwise 95\% confidence intervals are estimated by nonparametric bootstrap resampling across trials~\cite{Efron1979}.

\section{Results and discussion\label{sec:results}}

\subsection{Consistency check}
Synthetic cases were constructed using conditions informed by~\citet{Sanchez2026} to test recovery of prescribed material histories from radius data alone.
The maximum and equilibrium radii were $R_{\max}\approx\SI{229}{\micro\meter}$ and $R_{\mathrm{eq}}\approx\SI{38}{\micro\meter}$, with initial properties $G=\SI{15}{\kilo\pascal}$ and $\mu=\SI{0.1}{\pascal\cdot\second}$.
Each property was either held constant or reduced to half its initial value ($r_\theta^{\mathrm{true}}=0.50$) using a sigmoid transition centered at the first collapse, first rebound maximum, or second collapse.
The four choices for each property yielded $16$ history pairs, including the constant-property case.

The errors are sufficiently low for the application of the approach to experimental radius histories.
Using the constant material properties, the baseline fitting accuracy is $E^{\text{glob}} = 6.15\times10^{-8}$.
For the 15 cases in which one or both properties evolved, the mean $E^{\text{glob}}$ decreased from $(5.11\pm1.85)\times10^{-2}$ for the constant-property fits to $(1.03\pm0.92)\times10^{-4}$.
The mean relative material-history recovery errors were $(2.5\pm2.9)\times10^{-3}$ for $G(t)$ and $(3.3\pm2.7)\times10^{-3}$ for $\mu(t)$, with maximum errors of $9.5\times10^{-3}$ and $9.7\times10^{-3}$, respectively.

\subsection{Experimental comparisons}
\Cref{tab:experimental_summary} tabulates the $E^{\text{glob}}$ and the initial and final-to-initial ratios of the recovered properties for the ten experimental datasets.
Across datasets, mean $E^{\text{glob}}$ values range from $0.056$ to $0.076$ for constant-property fits and decrease to $0.013$ to $0.05$ for recovered histories.
The recovered errors indicate close agreement with the experimental radius histories.

\begin{figure*}[!t]
    \centering
    \includegraphics[width=\textwidth]{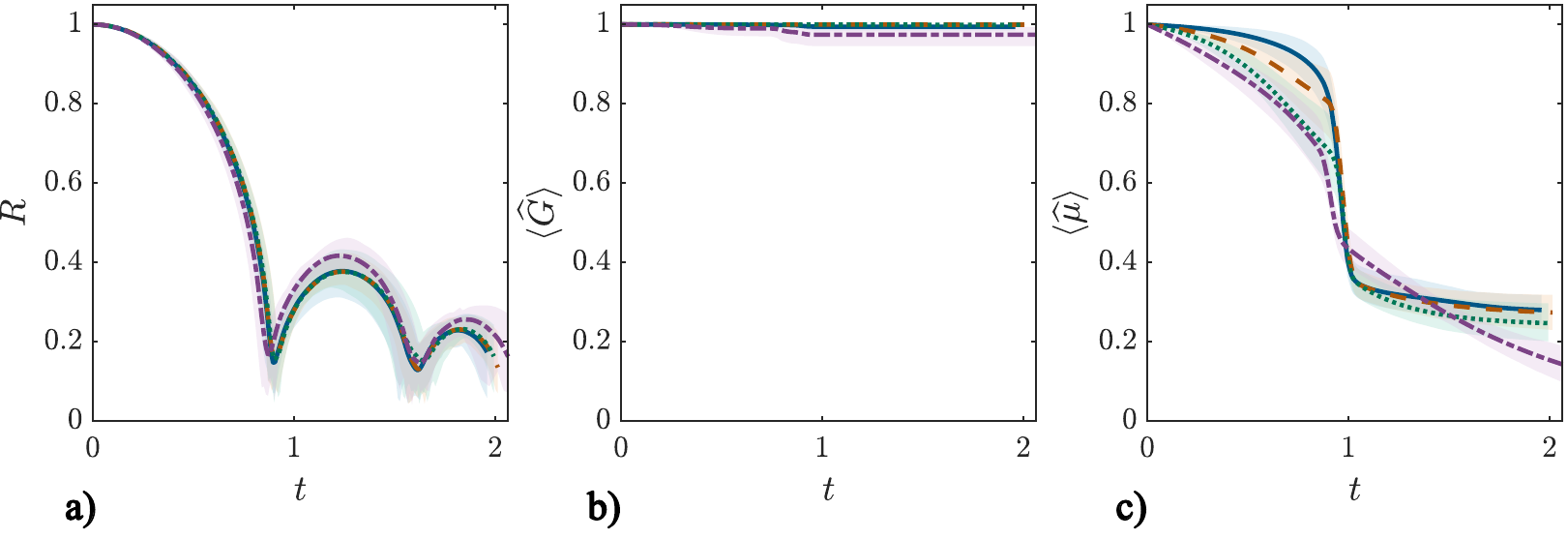}
    \caption{PA results: (a) nondimensional radius, (b) normalized shear modulus, $\langle\widehat{G}\rangle$, and (c) normalized viscosity, $\langle\widehat{\mu}\rangle$.
    Lines show dataset means: PA-3-015 (blue solid), PA-3-0225 (orange dashed), PA-3-030 (green dotted), and PA-5-030 (purple dash-dotted).
    Shading denotes the full experimental range in (a) and pointwise $95$\% bootstrap confidence intervals in (b,c).}
    \label{fig:PA_results}
\end{figure*}

An effectively constant shear modulus is recovered across the four PA datasets, while viscosity decreases near the first bubble collapse and rebound (see \Cref{fig:PA_results}).
For PA-3-015, the mean viscosity degradation is step-like near the first collapse.
As bisacrylamide concentration increases (i.e., PA-3-0225 and PA-3-030), the degradation occurs over a longer time interval.
For a fixed $0.3$~wt\% bisacrylamide, the viscosity degradation is more prolonged and larger for $5$~wt\% acrylamide than for $3$~wt\%, with mean relative degradations of $0.869$ and $0.754$, respectively.
While the approximately constant recovered shear modulus does not support widespread loss of network elasticity~\cite{Rubinstein2003,Calvet2004}, localized damage may still be taking place.
The collapse-associated viscosity degradation is consistent with the apparent viscosity changes inferred in polyacrylamide gels by \citet{Spratt2021}.
Its timing near the first collapse, where $|\dot{\varepsilon}|$ peaks, is compatible with a rate-dependent viscous response but cannot distinguish rate dependence from changes associated with prior loading.

\begin{table*}[!t]
\centering
\caption{Recovered experimental mechanical properties and global relative radius errors.
Values are reported as mean $\pm$ standard deviation across trials within each dataset.
Constant-property fits use $r_G=r_\mu=1$.}
\label{tab:experimental_summary}

\small
\setlength{\tabcolsep}{7pt}
\renewcommand{\arraystretch}{1.1}

\begin{tabular*}{\textwidth}{@{\extracolsep{\fill}}lcccccc@{}}
\toprule
& \multicolumn{2}{c}{$E^{\mathrm{glob}}$}
& \multicolumn{2}{c}{Initial properties}
& \multicolumn{2}{c}{Final-to-initial ratios} \\
\cmidrule(lr){2-3}
\cmidrule(lr){4-5}
\cmidrule(lr){6-7}
Dataset & Constant & Recovered
& $a_G$ [\si{\kilo\pascal}]
& $a_\mu$ [\si{\pascal\cdot\second}]
& $r_G$ & $r_\mu$ \\
\midrule
PA-3-015  & $0.057\pm0.006$ & $0.041\pm0.007$ & $16.14\pm2.28$ & $0.486\pm0.179$ & $0.994\pm0.025$ & $0.278\pm0.117$ \\
PA-3-0225 & $0.056\pm0.006$ & $0.043\pm0.007$ & $16.26\pm2.29$ & $0.601\pm0.225$ & $0.999\pm0.004$ & $0.273\pm0.135$ \\
PA-3-030  & $0.057\pm0.005$ & $0.046\pm0.007$ & $17.67\pm2.66$ & $0.764\pm0.25$ & $0.999\pm0.004$ & $0.246\pm0.139$ \\
PA-5-030  & $0.062\pm0.006$ & $0.033\pm0.007$ & $19.49\pm2.41$ & $0.48\pm0.16$ & $0.974\pm0.068$ & $0.131\pm0.142$ \\
KG-5      & $0.066\pm0.006$ & $0.037\pm0.003$ & $13.35\pm2.42$ & $0.405\pm0.043$ & $0.437\pm0.102$ & $0.107\pm0.154$ \\
KG-10     & $0.069\pm0.011$ & $0.02\pm0.005$ & $18.33\pm4.21$ & $0.268\pm0.041$ & $0.386\pm0.14$ & $0.011\pm0.002$ \\
PG-5-UM   & $0.058\pm0.006$ & $0.026\pm0.006$ & $11.45\pm2.25$ & $0.272\pm0.058$ & $0.831\pm0.176$ & $0.034\pm0.092$ \\
PG-10-UM  & $0.076\pm0.008$ & $0.016\pm0.003$ & $26.45\pm4.54$ & $0.252\pm0.063$ & $0.363\pm0.095$ & $0.011\pm0.002$ \\
PG-5-UT  & $0.074\pm0.004$ & $0.05\pm0.011$ & $19.54\pm2.51$ & $0.55\pm0.085$ & $1.000\pm0.001$ & $0.217\pm0.024$ \\
PG-10-UT & $0.066\pm0.005$ & $0.013\pm0.002$ & $25.09\pm2.39$ & $0.26\pm0.051$ & $0.476\pm0.086$ & $0.012\pm0.002$ \\
\bottomrule
\end{tabular*}
\end{table*}

\begin{figure*}[!t]
    \centering
    \includegraphics[width=\textwidth]{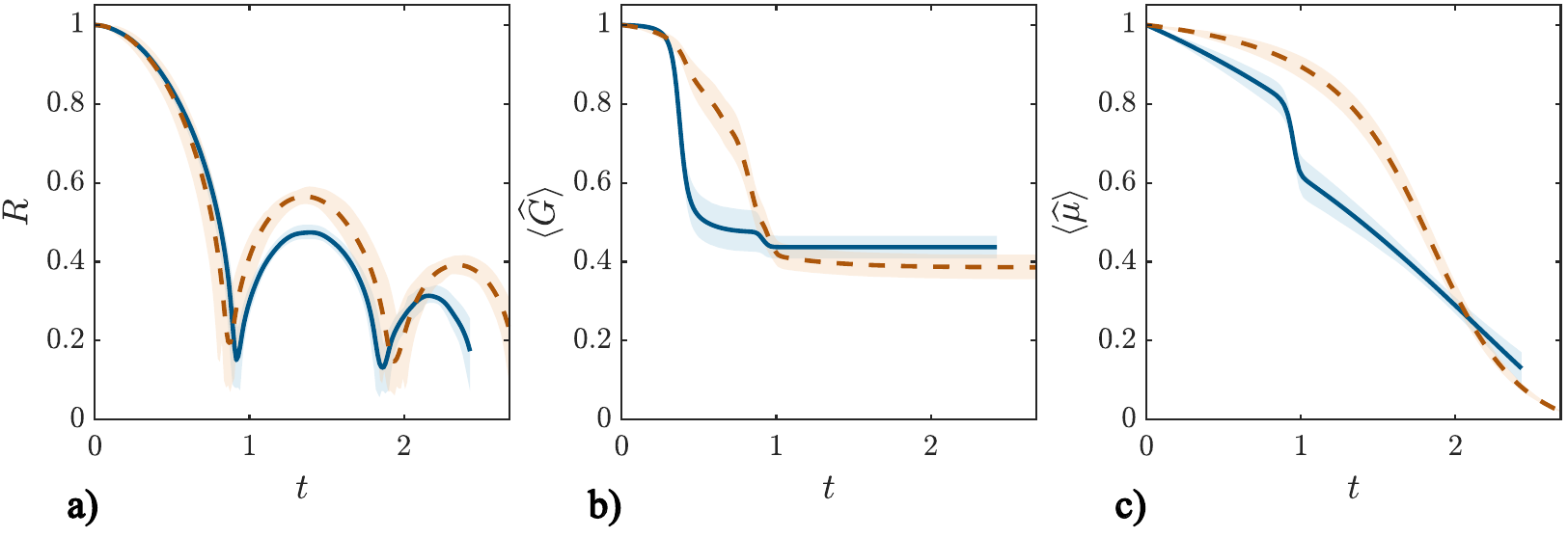}
    \caption{Knox gelatin results: (a) nondimensional radius, (b) normalized shear modulus, $\langle\widehat{G}\rangle$, and (c) normalized viscosity, $\langle\widehat{\mu}\rangle$.
    Lines show dataset means: KG-5 (blue solid) and KG-10 (orange dashed).
    Shading denotes the full experimental range in (a) and pointwise 95\% bootstrap confidence intervals in (b,c).}
    \label{fig:knox_results}
\end{figure*}

\Cref{fig:knox_results} shows both the recovered shear modulus and viscosity decrease in the Knox gelatin datasets and at different stages of the bubble dynamics.
Relative to KG-5, KG-10 has a higher mean initial shear modulus, which is consistent with \citet{Li2020}.
Additionally, KG-10 has lower mean initial viscosity, mean maximum radius, and mean maximum stretch ratio.
At a given stretch, the higher shear modulus further drives the collapse, while the lower viscosity is less dampening at a given deformation rate.
Both effects favor a faster initial contraction for KG-10 than KG-5 (see \cref{fig:knox_results}(a)).
The KG-5 shear modulus falls sharply after maximum expansion and gradually in KG-10, with comparable mean relative degradations of $0.563$ and $0.614$, respectively.
The viscosity degradation occurs before and during the first collapse in KG-5 and during the first rebound and second collapse in KG-10.
Temporally earlier KG-5 shear modulus changes are consistent with disruption of its less concentrated gelatin network during the initial loading.
The denser KG-10 network may sustain this loading before undergoing changes during later oscillations.
These results complement the temperature-dependent apparent property evolution reported by \citet{Chu2026} by showing how the timing of this evolution differs across gelatin concentrations.

Although the PA-5-030 and KG-5 trials reach similar mean maximum stretches of $7.90$ and $7.78$, their mean relative shear modulus degradations differ at $0.026$ and $0.563$, respectively.
This suggests that maximum stretch alone cannot explain the recovered shear modulus degradations and motivates consideration of differences in network structure.
Unlike the covalently crosslinked PA network, gelatin has physical triple-helix junctions, and fracture studies identify junction yielding and chain pullout as potential failure mechanisms~\cite{Baumberger2006,Frieberg2018}.
The shear modulus degradation suggests disruption of these junctions during the preceding expansion that becomes apparent during collapse. 

\begin{figure*}[!t]
    \centering
    \includegraphics[width=\textwidth]{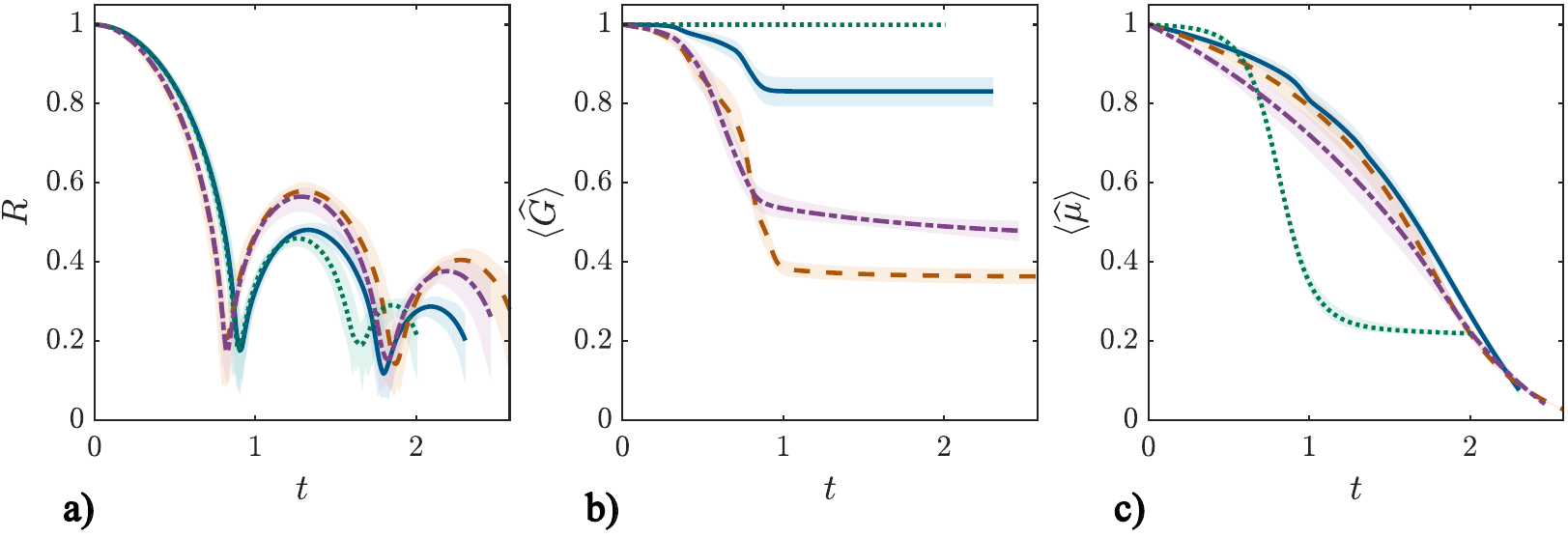}
    \caption{Porcine gelatin results: (a) nondimensional radius, (b) normalized shear modulus, $\langle\widehat{G}\rangle$, and (c) normalized viscosity, $\langle\widehat{\mu}\rangle$.
    Lines show dataset means: PG-5-UM (blue solid), PG-10-UM (orange dashed), PG-5-UT (green dotted), and PG-10-UT (purple dash-dotted).
    Shading denotes the full experimental range in (a) and pointwise 95\% bootstrap confidence intervals in (b,c).}
    \label{fig:porcine_results}
\end{figure*}

\Cref{fig:porcine_results} shows distinct fitted responses for the nominal 5\% porcine gelatin formulations.
The PG-5-UT shear modulus is approximately constant and PG-5-UM has a mean relative degradation of $0.169$.
As discussed for Knox gelatin, yielding of physical triple-helix junctions could weaken the load-bearing network and lead to a lower shear modulus in PG-5-UM.
Moreover, the PG-5-UM viscosity degradation is later and larger than in PG-5-UT.
In PG-5-UT, the early change in elastic resistance also leads to a sharp viscosity drop near the first collapse.
This drop reflects a nonlinear viscous response at high strain rates.
Both nominal 10\% formulations have higher mean initial shear moduli and larger relative shear modulus degradations than their nominal 5\% counterparts.
The differences in gelatin content and thermal-history effects on triple-helix formation~\cite{Djabourov1988} (see~\Cref{sec:experiments}) and boundary effects may contribute to the differences in the radial dynamics between the UM and UT datasets.
The glass bottom in PG-5-UT was about $2.2R_{\max}$ from the bubble center and can also alter the bubble dynamics and, in turn, the IMR-inferred material properties~\cite{kolluri2025determination}.

\subsection{Outlook}
The recovered material-property histories can guide the development of constitutive models for material evolution during cavitation.
For example, damage models can capture the distinct transition times by coupling internal variables differently to elastic resistance and viscous damping~\cite{Kachanov1958,Lemaitre1996,SimoJu1987}.
Stress-softening models~\cite{Mullins1969,Diani2009} and phase-field fracture formulations~\cite{Miehe2010} are other descriptions to evaluate.
Using these histories for constitutive-model development requires assessing their uncertainty and sensitivity to modeling assumptions.
Uncertainty in transition timing and magnitude could be quantified through ensemble-based data assimilation~\cite{Spratt2021,Evensen2009} or particle-based inference~\cite{Gordon1993,Doucet2001,Arulampalam2002,Kantas2015}.
Simulations with nonspherical dynamics~\cite{Yang2021Instability,Remillard2026,Remillard2026Instability} or spatially varying properties could test whether the histories compensate for omitted physics.
Independent measurements could then test whether the inferred degradations persist after cavitation.
Evidence of lasting softening or fracture would support an interpretation of irreversible damage~\cite{Hasan2021,Movahed2016,Mancia2019,RaayaiArdakani2019,Wang2026}.
Spatially resolved deformation measurements could link these changes to local stress and strain-rate fields~\cite{mcghee2023high,Mancia2021}.

\section{Conclusions\label{sec:conclusions}}
We present an IMR technique to find effective time-varying material properties from inertial cavitation bubble dynamics.
Shear-modulus and viscosity histories are parameterized using normalized sigmoid functions that include constant behavior and approach step-like or linear histories in limiting cases.
Synthetic tests recovered the tested histories from radius data for experimentally relevant conditions.
Allowing property evolution reduced the mean global relative radius error across the 502 PA and gelatin experimental trials within the Kelvin--Voigt framework.
Viscosity decreased across all the datasets, while the extent of shear modulus degradation depended on the material and its composition.
When both properties decreased, their degradations could occur at different stages of the bubble dynamics.
These findings show that characterizing effective mechanical property degradation requires resolving both its magnitude and timing in the elastic and viscous responses during high-strain-rate loading.

\section{Acknowledgments}
MRJ acknowledges support from the U.S. Department of Defense under the DEPSCoR program Award No. FA9550-23-1-0485 (PM Dr. Timothy Bentley).
JBE, MRJ, and JY acknowledge support from the U.S. National Science Foundation (NSF) under Grant Nos. 2232426, 2232427, and 2232428, respectively.
VS acknowledges support from the National Science Foundation (NSF) Graduate Research Fellowship Program.
SW and JY acknowledge Yujie Zhang and Aidan Landis for their input on data post-processing.
Funding agencies were not involved in study design; in the collection, analysis and interpretation of data; in the writing of the report; or in the decision to submit the article for publication.
The opinions, findings, and conclusions, or recommendations expressed are those of the authors and do not necessarily reflect the views of the funding agencies.
OpenAI’s ChatGPT was used to assist in coding, postprocessing visualizations, and all outputs were verified multiple times by the authors. 

\appendix

\bibliographystyle{elsarticle-num-names}
\bibliography{ref}

\end{document}